\documentclass[aps,prl,twocolumn,superscriptaddress,groupedaddress]{revtex4}  % for review and submission
\usepackage{epsfig}
\usepackage{natbib}
\usepackage{color}
\graphicspath{{./Figures/}}
\usepackage{amsmath}
\usepackage{amssymb}
\usepackage{titlesec}

\titlespacing\section{0pt}{12pt plus 4pt minus 2pt}{0pt plus 2pt minus 2pt}
\titlespacing\subsection{0pt}{12pt plus 4pt minus 2pt}{0pt plus 2pt minus 2pt}
\titlespacing\subsubsection{0pt}{12pt plus 4pt minus 2pt}{0pt plus 2pt minus 2pt}

\begin{document}
\newcommand{\h}{\phi}
\newcommand{\mon}{\begin{displaymath}}
\newcommand{\moff}{\end{displaymath}}
\newcommand{\sumi}[1]{\sum_{{#1}=-\infty}^{\infty}}
\renewcommand{\b}[1]{\mbox{\boldmath ${#1}$}}
\newcommand{\sumy}{\sum_{\b{y}}}
\newcommand{\sumz}{\sum_{\b{z}}}
\newcommand{\pd}[2]{\frac{\partial {#1}}{\partial {#2}}}
\newcommand{\od}[2]{\frac{d {#1}}{d {#2}}}
\newcommand{\inti}{\int_{-\infty}^{\infty}}
\newcommand{\eon}{\begin{equation}}
\newcommand{\eoff}{\end{equation}}
\newcommand{\eaon}{\begin{eqnarray}}
\newcommand{\eaoff}{\end{eqnarray}}
\newcommand{\e}[1]{\times 10^{#1}}
\newcommand{\chem}[2]{{}^{#2} \mathrm{#1}}
\newcommand{\s}{s}
\newcommand{\zetaexp}{\left( \zeta e^{q \s t} \right)}
\newcommand{\taunuc}{\tau_{nuc}}
\newcommand{\eq}[1]{Eq. (\ref{#1})}
\newcommand{\ev}[1]{\left\langle #1 \right\rangle}
\newcommand{\mat}[1]{\bf{\mathcal{#1}}}
\newcommand{\fig}[1]{Fig. \ref{#1}}
\newcommand{\ub}{U_b}
\newcommand{\dcoal}[2]{D_{#1#2}}
\newcommand{\ph}{\dcoal{H}{1}}
\newcommand{\Prob}{\mathrm{Pr}}
\renewcommand{\H}{H}
\newcommand{\Co}{{\cal{C}}}
\newcommand{\Gam}{\Gamma}
\newcommand{\la}{\langle}
\newcommand{\ra}{\rangle}
\newcommand{\Oh}{{\cal O}}
\newcommand{\pib}[2]{\pi^b_{#1, #2}}
\newcommand{\un}{U_n}
\newcommand{\param}{\Gamma}
\newcommand{\moo}{\mu_{i}}
\newcommand{\mot}{\mu_{v}}
\newcommand{\mto}{\mu_{u}}
\newcommand{\soo}{s_{i}}
\newcommand{\doo}{\delta_{i}}
\newcommand{\sot}{s_{v}}
\newcommand{\sto}{s_{u}}
\newcommand{\sooe}{s_{i,\text{eff}}}
\newcommand{\dooe}{\delta_{i,\text{eff}}}
\newcommand{\sote}{s_{v,\text{eff}}}
\newcommand{\stoe}{s_{u,\text{eff}}}
\newcommand{\lot}{\lambda_{v}}
\newcommand{\lto}{\lambda_{u}}
\newcommand{\lott}{\Lambda_{v}}
\newcommand{\ltot}{\Lambda_{u}}
\newcommand{\Tot}{T_{v}}
\newcommand{\Tto}{T_{u}}
\newcommand{\tot}{t_{v}}
\newcommand{\tto}{t_{u}}
\newcommand{\rzzoo}{r_{w i}}
\newcommand{\roozz}{r_{i w}}
\newcommand{\rooot}{r_{i v}}
\newcommand{\rzzto}{r_{w u}}
\newcommand{\pot}{p_{i}}
\newcommand{\pto}{\pi(s_{u})}
\newcommand{\ptt}{\pi(s_{v})}
\newcommand{\pv}{p_v}
\newcommand{\ps}{\pi(s)}
\newcommand{\fto}{f_{u}}
\newcommand{\foo}{f_{i}}
\newcommand{\pcross}{P_{\text{cross}}}
\newcommand{\tsw}{\tau_{nm}}
\newcommand{\tmrca}{J}
\newcommand{\appropto}{\mathrel{\vcenter{
  \offinterlineskip\halign{\hfil$##$\cr
    \propto\cr\noalign{\kern2pt}\sim\cr\noalign{\kern-2pt}}}}}
\newcommand{\lp}{\left(}
\newcommand{\rp}{\right)}

\title{The competition between simple and complex evolutionary trajectories in asexual populations}
    % D0 authors (remove the first 3 lines
                             % of this file prior to submission, they
                             % contain a time stamp for the authorlist)
                             % (includes institutions and visitors)
                             
\author{I. E. Ochs}
\affiliation{Department of Physics, Department of Organismic and Evolutionary Biology, and Center for Systems Biology, Harvard University, Cambridge, Massachusetts 02138}

\author{Michael M. Desai}
\affiliation{Department of Physics, Department of Organismic and Evolutionary Biology, and Center for Systems Biology, Harvard University, Cambridge, Massachusetts 02138}
                             
\date{\today}

\begin{abstract}

On rugged fitness landscapes where sign epistasis is common, adaptation can often involve either individually beneficial ``uphill'' mutations or more complex mutational trajectories involving fitness valleys or plateaus. The dynamics of the evolutionary process determine the probability that evolution will take any specific path among a variety of competing possible trajectories. Understanding this evolutionary choice is essential if we are to understand the outcomes and predictability of adaptation on rugged landscapes.
We present a simple model to analyze the probability that evolution will eschew immediately uphill paths in favor of crossing fitness valleys or plateaus that lead to higher fitness but less accessible genotypes.
We calculate how this probability depends on the population size, mutation rates, and relevant selection pressures, and compare our analytical results to Wright-Fisher simulations.
We find that the probability of valley crossing depends nonmonotonically on population size: intermediate size populations are most likely to follow a ``greedy'' strategy of acquiring immediately beneficial mutations even if they lead to evolutionary dead ends, while larger and smaller populations are more likely to cross fitness valleys to reach distant advantageous genotypes. We explicitly identify the boundaries between these different regimes in terms of the relevant evolutionary parameters. Above a certain threshold population size, we show that the degree of evolutionary ``foresight'' depends only on a single simple combination of the relevant parameters.
\end{abstract}

\maketitle

\section*{Background}

In an adapting population, evolution often has the potential to follow many distinct mutational trajectories. In order to predict how the population will adapt, we must understand how evolution chooses among these possibilities. Many experimental and theoretical studies have analyzed this question, focusing primarily on the simple case where epistasis is absent, so that each mutation has some fixed fitness effect \cite{perfeito07, tenaillon12, lang13, gerrishlenski98, schiffels11, good12}. This work can explain the probability that a given mutation will fix as a population adapts, as a function of its fitness effect, the population size, mutation rate, distribution of fitness effects of other mutations, and other parameters of the evolutionary process.

However, the fitness effect of a mutation often depends on the genetic background in which it occurs. A particularly interesting form of this phenomenon, \emph{sign epistasis}, occurs when several mutations are individually neutral or deleterious but their combination is beneficial \cite{weinreich05a}. Sign epistasis has been observed repeatedly in experiments \cite{weinreich06, poelwijck07, Kvitek2011, Silva2011, Dawid2010, salverda2011initial}, and plays a central role in the evolution of complex phenotypes that involve multiple interacting components. When sign epistasis is present, adaptation can involve passing through genotypes of lower fitness --- i.e. a population may have to cross a fitness valley or plateau. Thus the fate of a mutation depends not only on its fitness, but also on its adaptive potential \cite{woods2011second}.

Several recent theoretical studies have analyzed the evolutionary dynamics of fitness valley crossing \cite{weinreich05b, Iwasa2004, Weissman2009, gokhale2009pace}. This work has focused on calculating the rate at which adapting populations cross a valley or plateau, in the absence of any other possible mutational trajectories. However, individually beneficial mutations may often compete with more complex evolutionary trajectories. We must then ask how likely evolution is to eschew the immediately uphill paths, and instead cross valleys or plateaus to reach better but less accessible genotypes. In other words, when the fitness landscape is rugged, we wish to understand whether evolution will take the more ``farsighted'' path to reach distant advantageous genotypes, rather than a ``greedy'' trajectory that fixes immediately beneficial mutations regardless of whether these may lead to evolutionary dead ends.

In this article, we analyze this evolutionary choice between immediately beneficial mutations and more complex mutational trajectories that ultimately lead to higher fitness. We calculate the probability that an adapting population will follow each type of competing trajectory, as a function of the population size, mutation rates, and selection pressures. We focus on asexual populations, where the only way for a population to acquire a complex adaptation is for a single lineage to acquire each mutation in turn. Our analysis is similar in spirit to earlier work which also considered the tradeoff between short-term and long-term fitness advantages \cite{Rozen2008, Handel2009, Jain2011, van2000metastable}. However, these earlier studies dealt with competition between different strictly uphill or neutral paths, and considered the case where the less beneficial initial mutation led to better long-term evolutionary opportunities. In contrast, our analysis describes the competition between uphill mutations and more complex trajectories. While these two cases can be qualitatively similar in very small populations, they lead to very different dynamics in larger populations where the sign of the effect of the intermediate mutation can play a crucial role.

Our results show that population size has a crucial impact on how ``farsighted'' evolution can be. This dependence is not monotonic: evolution at intermediate population sizes is most ``greedy,'' while both larger and smaller populations are more likely to eschew uphill paths in favor of complex trajectories. In large populations, our results show that a single parameter reliably predicts the extent of this evolutionary ``foresight'' across a wide range of parameters. Finally, we describe how our analysis can be generalized to predict how evolution will choose among even more complex trajectories, such as broad fitness valleys with multiple intermediate genotypes, and we discuss evolution in genotype spaces with many possible evolutionary paths.

%----------------------------------------------------------------------------------------------------------------------------------
\section*{Methods}

We are interested in how a population makes an evolutionary choice when confronted with multiple possible mutational trajectories. Specifically, we focus on the extent to which adaptation proceeds by crossing fitness valleys rather than acquiring immediately beneficial (uphill) mutations. Of course, the relative frequency of valley crossing will depend on the number of available fitness valleys, their depth, and the fitness advantage of the multiple-mutants, as well as the distribution of fitness effects (DFE) of the uphill mutants. Our goal is to understand how the prevalence of valley crossing depends on these factors.

\subsubsection*{Model}
\begin{figure}[t]
  \includegraphics[width=.95\linewidth]{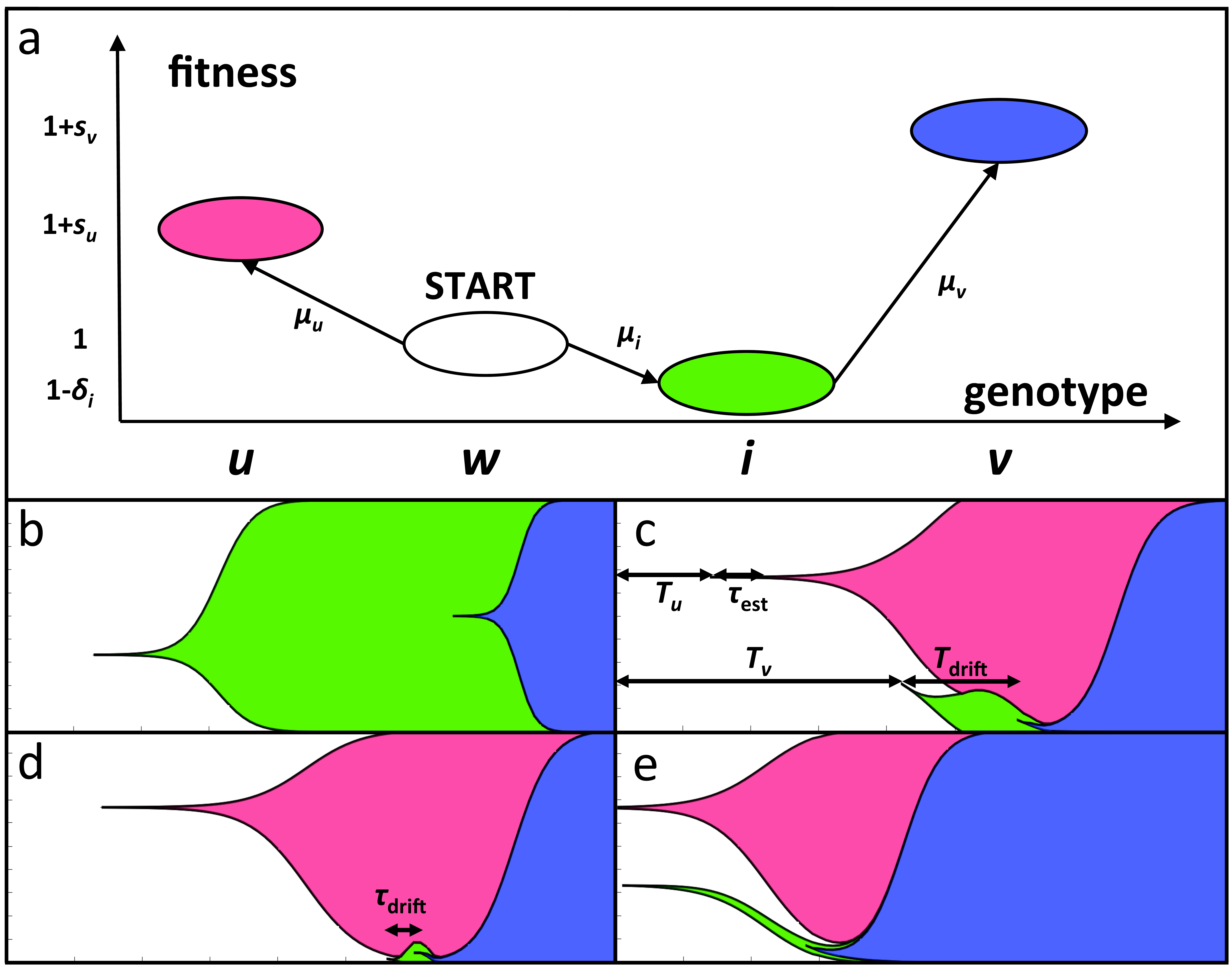}
  \caption{\textbf{Model and characteristic trajectories.}
      (a) The model to study fitness valley crossing prevalence. The population starts as wild type ($w$), and then acquires uphill mutations ($u$) at rate $\mto$ that confer an immediate fitness advantage $\sto$, and acquires deleterious fitness valley intermediates ($i$) at rate $\moo$ with fitness deficit $\doo$ on which background double-mutants ($v$) with fitness $\sot>\sto$ arise at rate $\mot$. (b)-(e) The four main forms of fitness valley crossing. (b) Small populations are characterized by low genetic diversity and strong genetic drift, leading \textbf{sequential fixation} of intermediates to dominate the dynamics. (c) For larger populations, genetic diversity is maintained longer, and double mutants will tend to arise on transient single-mutant backgrounds, a process known as \textbf{stochastic tunneling}. (d) If the drift time is small compared to the maximal rate of change in background fitness, we can approximate the drift time of the intermediate by its expectation, dramatically simplifying the mathematical analysis. (e) For very large populations, we can treat single-mutants deterministically, in a process dubbed \textbf{semi-deterministic tunneling}.}
      \label{fig:muller}
      \end{figure}

Throughout most of this article, we consider the simplest context in which we can address this question: the choice between a single uphill path and a single fitness valley. Specifically, we consider a haploid asexual population of constant size $N$ which can either acquire an uphill mutation ($u$) that confers an immediate fitness advantage $\sto$, or alternatively acquire a deleterious fitness valley intermediate ($i$) with fitness deficit $\doo$ on which background a double-mutant ($v$) with fitness $\sot > \sto$ can arise. This scenario is illustrated in \fig{fig:muller}. We also consider the case of a fitness plateau, where $\doo = 0$.

Because we are interested in the evolutionary choice between competing mutational trajectories, we assume that these two trajectories are mutually exclusive, so that only one genotype (either $u$ or $v$) can eventually fix in the population. As a measure of evolutionary foresight, we analyze the probability that the double-mutant $v$ fixes as a function of the relevant mutation rates, selection coefficients, and population size. In some situations, we could imagine that after either genotype $u$ or $v$ fixes, another set of competing potential trajectories become available. In this case, our analysis predicts the long-term relative ratio of fixed uphill versus valley-crossing mutations. In the Discussion, we consider how this model can be extended to the situation where there are many different competing uphill paths and valleys, and to broader fitness valleys involving multiple intermediate genotypes.

\subsubsection*{Simulations}

In addition, we compare our analytical predictions for valley crossing probability to Wright-Fisher simulations. Each simulated population was evolved until either the uphill genotype or valley-crossing genotype fixed. Valley crossing probabilities were then inferred from the number of trials in which the valley-crossing genotype fixed, out of 1000 trials per parameter set.

\section*{Results}

In the absence of the uphill genotype, fitness valley crossing can be modeled as a homogeneous Poisson process with rates as calculated by \cite{Weissman2009}. In small populations, the primary role of the uphill genotype is to introduce an effective \emph{time limit} on this process: once an uphill mutation destined to survive drift first occurs, it very quickly fixes, leading to the extinction of the wild-type. The probability of valley-crossing can thus be calculated as the probability that the intermediate $i$ fixes before the uphill genotype $u$. An example of this is shown in \fig{fig:muller}b.

In larger populations, the dynamics are more complex, as illustrated in \fig{fig:muller}c. Rather than leading to a single cutoff time for valley-crossing to occur, the single-mutant occurs and gradually increases in frequency. This leads to a decline in the size of the wild-type background on which intermediate and valley-crossing mutants can arise, and a corresponding increase in the mean fitness of the population (Fig. \ref{fig:muller}c). These effects gradually reduce the rate at which intermediates are produced, and make these intermediates effectively more deleterious relative to the mean fitness. These factors reduce the rate of the valley-crossing process. Thus valley-crossing becomes an \emph{inhomogenous} Poisson process, with a rate that depends on the random appearance time $T_u$ of the uphill mutant.

In general, these effects of interference and tunneling are complex. However, the analysis becomes simpler in two specific regimes. When the expected drift time of the intermediate genotype is short, we can neglect the changing background fitness due to the uphill mutant during this drift time (Fig. \ref{fig:muller}d). Alternatively, for very large populations ($N \mu>1$), the Poisson process approximation breaks down and both uphill and intermediate mutations can be treated deterministically (Figure \ref{fig:muller}e), and only the valley-crossing genotype must be treated stochastically.

\begin{figure}[b]
  \includegraphics[width=1\linewidth]{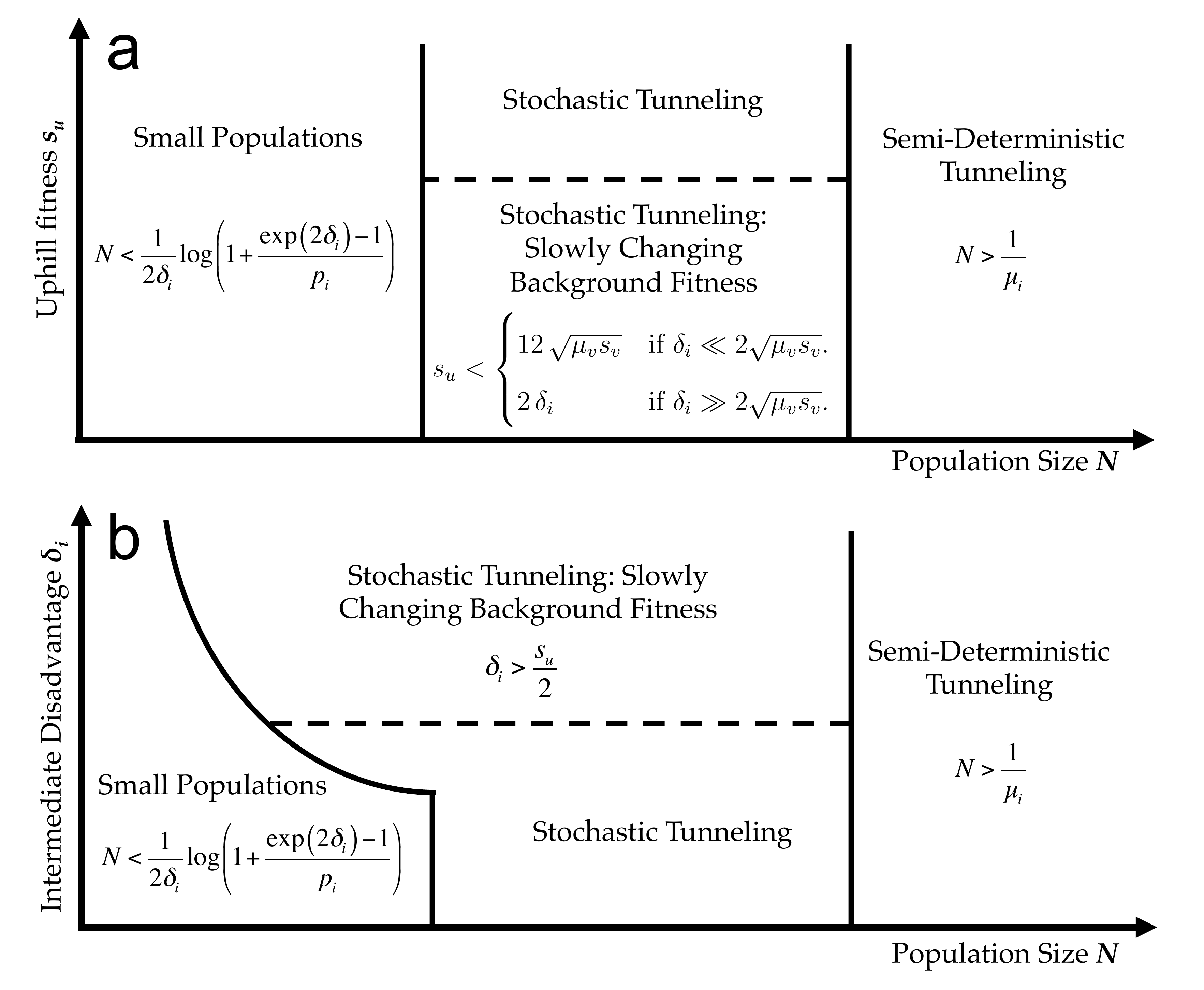}
  \caption{\textbf{Regimes of valley crossing.}
      Phase plot summarizing the different regimes of fitness valley crossing.}
   \label{fig:phase}
      \end{figure}

These various regimes are illustrated in \fig{fig:phase}. We now analyze each in turn, assuming weak selection ($s_j<1$) for all genotypes throughout. Taken together, this provides a complete picture of the probability that evolution will eschew the immediately uphill path in favor of the more complex adaptation.

\subsubsection*{Small Populations}

When the population size is small enough that the probability of stochastic tunneling is very low, the population is generally clonal or nearly clonal, and moves in Markovian jumps between neighboring genotypes. The transition between genotypes $i$ and $j$ occurs at rate \eon r_{i j} = N \mu_{i j} \pi_{i j} , \label{eq:ab_rate} \eoff where $\pi_{i j}$ is the probability that a single $j$ mutant will give rise to a lineage that fixes, given by the standard formula, \eon \pi_{i j} = \frac{1-e^{-2(s_{j}-s_{i})}}{1-e^{-2N(s_{j}-s_{i})}}. \eoff We refer to this as the sequential fixation regime. Because we are considering neutral and weakly deleterious intermediates, we account for the possibility of back-mutation to the wild type if the intermediate fixes. Therefore, the process can be modeled as an absorbing states Markov chain, where the wild type and intermediates act as transient states, and the uphill and double-mutant genotypes act as absorbing states. From elementary Markov chain theory, we find \begin{align} \pcross &= \frac{\rzzoo  \rooot}{\rzzoo  \rooot + \rzzto (\roozz + \rooot)}\notag\\ 
&= \left[1 + \left( \frac{\pi_{wu}}{\pi_{wi}} \right)\left( \frac{\mto}{\moo} \right) \left(1+\frac{\moo \pi_{iw}}{\mot \pi_{iv}} \right)\right]^{-1}. 
\end{align} As the population size increases, $\pi_{wu} \rightarrow 2 \sto$ and $\pi_{wi}\rightarrow 0$, so $\frac{\pi_{wu}}{\pi_{wi}} \rightarrow \infty$, and $\pcross \rightarrow 0$. Thus we find that within the sequential fixation regime, larger population sizes are less likely to cross fitness valleys.

\subsubsection*{Stochastic Tunneling}

For large populations, the probability that deleterious intermediates will fix declines drastically, and successful double mutants will instead arise on the unfixed single-mutant background in a process known as \emph{stochastic tunneling} \cite{Iwasa2004}. This transition occurs when
\eon N > \frac{1}{2 \delta_i} \log \left[ 1 + \frac{\exp (2 \delta_i) - 1}{p_v} \right], \eoff
where $p_v$ is the probability that the intermediate lineage survives drift long enough to give rise to an ultimately successful double mutant lineage (we will explicitly calculate this probability below).
We can then model the appearance of a intermediate mutant lineage destined to give rise to a double mutant lineage as a Poisson process. The rate at $\lot$ at which these lineages appear is given by the rate at which intermediate mutations arise times the probability of success of the lineage, integrated over the drift time $t_d$ after appearance of the single-mutant intermediate:
\eon \lot=N_\text{wt} \moo \int_0^\infty \frac{\partial \pot(t_d)}{\partial t_d} dt_d. \eoff
Here $N_\text{wt}$ is the wild-type population size, and $\pot (t_d)$ is the cumulative probability that a single-mutant lineage will give rise to a successful double-mutant lineage by time $t_d$ after it appears. This probability is given by \cite{Weissman2009}:
\begin{footnotesize}
\eon \pot (t_d) = 2 \left( \frac{(a_+ -1)(1-a_{-})(1-e^{-(1-\dooe)(a_{+}-a_{-})t_d})}{a_+ -1+(1-a_{-})(1-e^{-(1-\dooe)(a_{+}-a_{-})t_d})} \right),\eoff \end{footnotesize}
where
\begin{footnotesize} \eon a_{\pm} = \frac{2 -\dooe - \mot \sote \pm \sqrt{(\dooe+\mot \sote)^2+4\mot \sote}}{2(1-\dooe)}, \eoff\end{footnotesize}
and $\delta_{i,\text{eff}}$ and $s_{v,\text{eff}}$ are the fitnesses of the intermediate and valley-crossing genotypes relative to the (time-dependent) mean population fitness. These effective fitnesses and $N_\text{wt}$ depend on the background at time $t_d$ in a way we must now consider.
The background in turn is determined by the frequency $\fto (\Tot + t_d)$ of the uphill genotype at time $t_d$ after the appearance of the first single-mutant intermediate lineage destined for success at time $\Tot$ (Fig. \ref{fig:muller}c).
Thus the uphill genotype frequency sets the fitness background on which valley-crossing probabilities are determined.

To calculate the relevant effective parameters, we note that the appearance of uphill lineages destined for success can be modeled as a Poisson process. Moreover, because the valley-crossing genotypes make up a tiny fraction of the population (unless the double-mutant has already established), we can treat the genetic background on which these uphill lineages appear as essentially fixed. Therefore, the first uphill lineage destined to survive genetic drift will appear at time $\Tto$, distributed exponentially with rate
\eon \lto=N \mto \pi_{wu} \approx N \mto (2 \sto) . \eoff
Once a successful uphill lineage appears, we assume it establishes in time $\tau_\text{est}=\gamma_e/(2\sto)$, where $\gamma_e \approx .577$ is the Euler-Mascheroni constant \cite{Desai2007} , and then sweeps deterministically according to
\begin{equation}
\fto(\hat{t}) = \frac{1-\exp\left[-(\mto+\sto)\hat{t}\,\right]}{1+(\sto/\mto)\exp\left[-(\mto+\sto)\hat{t}\,\right]}, \label{eq:fto}
\end{equation}
where $\hat{t}\equiv t-\Tto-\tau_\text{est}$ is the time after establishment of the uphill mutant.

Conditioning on the appearance time $\Tto$, we can thus work out our effective parameters \eaon N_\text{wt}(t \, | \, \Tto) & = & N(1-\fto(t-\Tto-\tau_\text{est})) \\ s_{k,\text{eff}}(t+t_d \, | \, \Tto) & = & s_k-\fto(t+t_d-\Tto-\tau_\text{est}) \sto . \eaoff These encompass the two main effects of the sweeping uphill mutation on the valley crossing probability: the first represents the declining wild-type background on which new mutations can arise, and the second represents the decreasing relative fitness of the valley-crossing lineage.

We are interested in the probability that a double-mutant lineage destined for success appears before the uphill genotype fixes. Integrating over all possible appearance times $\Tto$, this is given by: 
\begin{scriptsize} \begin{align} \pcross = & \int_0^{\infty} dt_u \left(\lto e^{-\lto t_u} \right) \times \label{eq:full_p} \\ & \left[ 1 - \exp\left[-\int_0^{\infty} dt \left(N_\text{wt}(t,\tto) \moo \int_0^\infty dt_d \frac{\partial \pot(t+t_d,t_u)}{\partial t_d} \right)\right]\right] . \nonumber \end{align} \end{scriptsize} This integral is a complete solution for the probability of valley-crossing in the stochastic tunneling regime, provided that the population size is small enough that the Poisson process approximation above holds. Although it does not have a simple closed-form solution, we can easily evaluate the integral numerically. Alternatively, there is a simple and relevant parameter regime in which background fitnesses change slowly. We now consider this case, and show that it allows us to evaluate our expression for the valley-crossing probability explicitly. In a later section below, we turn to the alternative case where the Poisson process approximation breaks down, and we can instead treat all single-mutants deterministically; the valley-crossing probability also simplifies considerably in this very large population regime.

\subsubsection*{Slowly Changing Background Fitness}

One of the main complications of \eq{eq:full_p} is the integral over possible drift times $t_d$, which reflects the increasing effective deleteriousness of the intermediate genotype as the uphill genotype sweeps to fixation and increases the mean fitness of the background population. However, when $\sto$ is small or $\doo$ is large, this background fitness changes slowly compared to the intermediate drift time. In this case, we can treat the background during intermediate drift as effectively constant (Fig. \ref{fig:muller}d). This eliminates the need for an integral over $t_d$, since the time-dependent probability of success of a single-mutant at time $t$ is fully determined by $\fto(t+t_d)\approx \fto(t)$.

We can further simplify the analysis if we treat the probability of crossing the valley as a function of two probabilities: the probability $P_{v,1}$ that the first successful valley-crossing lineage appears before the first successful uphill lineage establishes, and the probability $P_{v,2}$ that a successful valley-crossing lineage appears after the uphill mutant establishes:
\begin{equation}
	\pcross = P_{v,1} + \left(1-P_{v,1} \right) P_{v,2}.
\end{equation}

The calculation of $P_{v,1}$ takes place on a purely wild-type background, so we can use
\begin{equation}
	\lot = N\moo p_v ,
\end{equation}
where $p_v$, the probability that the intermediate lineage survives drift long enough to give rise to an ultimately successful double mutant lineage, is given by \cite{Weissman2009}:
\begin{align} p_v &= -\doo+\sqrt{\doo^2+4\mot \sot} \notag\\ &\approx \begin{cases}
		2\sqrt{\mot \sot} & \text{if } \doo \ll 2\sqrt{\mot \sot} \\
		2\mot \sot/\doo & \text{if } \doo \gg 2\sqrt{\mot \sot}.
	\end{cases} \label{eq:pi} \end{align}
Meanwhile, $\lto$ is unchanged from the original analysis. $P_{v,1}$ is determined by a race between these two exponential random variables.
Using basic properties of the exponential, the probability that the double-mutant appears before the uphill genotype establishes is therefore
\begin{align}
	P_{v,1} &= \begin{cases}
		\left(\frac{\lot}{\lot+\lto}\right) e^{-\lto \Delta \tau} &\text{if } \Delta \tau>0\\
		1-\left(\frac{\lto}{\lot+\lto}\right) e^{-\lot \Delta \tau} &\text{if } \Delta \tau<0 ,
	\end{cases} %\label{eq:psimp}
\end{align}
where $\Delta \tau$ represents the difference between the mean drift time $\tau_\text{drift}$ of the valley-crossing lineage and the mean establishment time $\tau_\text{est}$ of the uphill lineage,
\eon \Delta \tau = \tau_\text{drift} - \tau_\text{est} . \eoff
Here $\tau_\text{est}$ is as given above, and from \cite{Weissman2009}, we can approximate the drift time as
\begin{align}
	\tau_\text{drift} &= \begin{cases}
		\log 2/\sqrt{\mot \sot}  &\text{if } \doo \ll 2\sqrt{\mot \sot}\\
		1/\doo  &\text{if } \doo \gg 2\sqrt{\mot \sot}.
	\end{cases}  \label{eq:tdrift}
\end{align}

If the successful uphill lineage establishes, the first successful valley-crossing lineage still has a chance to appear and outcompete it, albeit on a declining wild-type background. Thus for $P_{v,2}$ we get a similar integral as in the original analysis. However, since we are approximating $\pot$ as constant, the rate of successful lineage generation simplifies to
\begin{align}
	\hat{\lambda}_v(t) &= \moo \, N_\text{wt} \; p_i (\delta_{i,\text{eff}},s_{v,\text{eff}})\notag \\
	&=  \moo \, N(1-\fto) \biggl(-(\doo+\sto \fto) \\
	&\quad +\sqrt{(\doo+\sto \fto)^2+4\mot (\sot-\sto \fto)}\biggr) \notag.
\end{align}
Integrating and assuming mutation rates are small compared to selection pressures, we find:
\begin{small}
\eon \int_0^\infty dt \hat{\lambda}_v(t) \approx \left(\frac{\log{N\sto}}{\sto}\right) N \moo \pot = \left(\frac{\log{N\sto}}{\sto}\right) \lot . \eoff \end{small}
Combining these results, we find
\begin{align}
	\pcross &= P_{v,1} + \left(1-P_{v,1}\right) \left(1-e^{-\left(\frac{\log{N\sto}}{\sto}\right) \lot}\right) \notag \\
	&= 1 -  \left(1-P_{v,1}\right) e^{-\left(\frac{\log{N\sto}}{\sto}\right) \lot} .
\end{align}
We expect this result to be valid provided that $\doo$ is effectively constant over the expected drift time. This will hold when
\begin{align}
	\sto &\ll \begin{cases}
		2 \sqrt{\frac{2}{\log 2}} \sqrt{\mot \sot} & \text{if } \doo \ll 2\sqrt{\mot \sot}\\
		2 \doo & \text{if } \doo \gg 2\sqrt{\mot \sot}
	\end{cases} . \label{eq:su}
\end{align}

\begin{figure*}[t]
  \includegraphics[width=.9\linewidth]{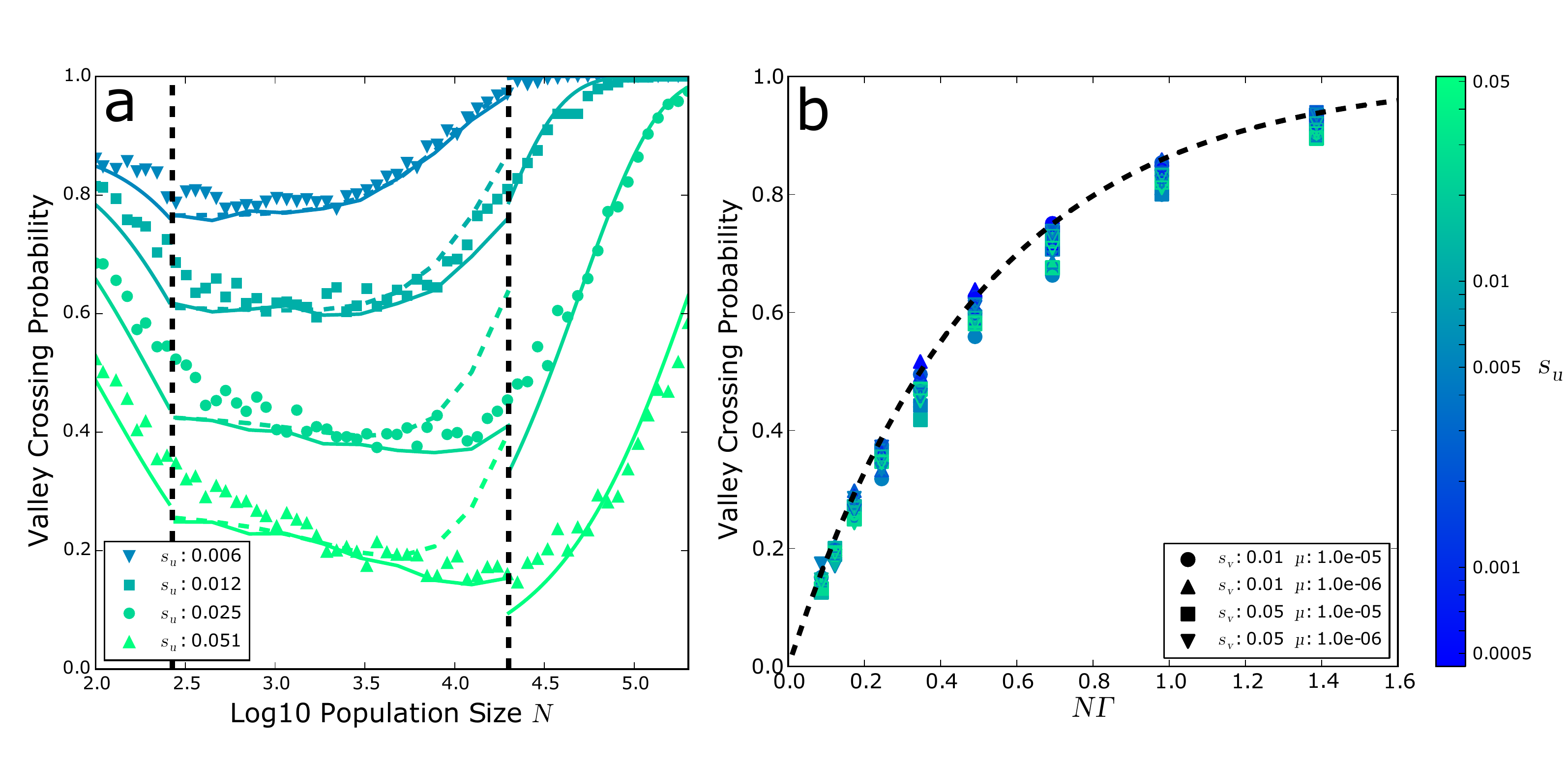}
  \caption{\textbf{Valley crossing probability.}
      (a) Simulation results for $\mto=5\times 10^{-6}$, $\moo=\mot=5 \times 10^{-5}$, $\doo=0$, and $\sot=.07$. The black vertical dashed lines indicate the boundaries between sequential fixation, stochastic tunneling, and semi-deterministic tunneling. Markers represent inferred valley-crossing probabilities from 1000 simulations per point. Lines represent theoretical predictions in each regime: in the stochastic tunneling regime, dashed lines represent the slowly changing background fitness approximation, and solid lines represent the full integral solution (\eq{eq:full_p}). The color of the line indicates the uphill fitness $\sto$. (b) Crossing probability for populations in the semi-deterministic regime across a wide range of parameters, plotted against the predictive parameter $N \param$. Filled markers represent deleterious intermediates ($\doo = 10\sqrt{\mot \sot}$), while open markers represent neutral intermediates ($\doo = 0$).}
   \label{fig:data}
      \end{figure*}

\subsubsection*{Semi-Deterministic Tunneling}

We now consider the case where $N\moo>1$ and $N\mto>1$, and hence the Poisson process approximation used to derive \eq{eq:full_p} breaks down. Fortunately, in this regime the number of single-mutant intermediates and uphill mutants in the population are well approximated by their deterministic expectation (Fig. \ref{fig:muller}e).
Thus the only random variable is the appearance time of the first successful double-mutant lineage, which occurs with rate
\eon \lot(t) = N \foo \mot \pi_{iv} \approx 2 N \foo \mot (\sot - \sto \fto) . \eoff
Because intermediates never make up a large portion of the population, $\fto$ is unaffected by $\foo$, and hence is still given by \eq{eq:fto}. We can then approximate the frequency $\foo$ of the single-mutant intermediates using mutation-selection balance with a declining wild-type population:
\eon \foo = \foo^* (1 - \fto) , \eoff
where $\foo^*$ gives the independent deterministic dynamics (mutation-selection balance) of single mutants on the wild-type background, and $(1 - \fto)$ is the size of the wild-type background.
It is useful to transform this into an integral in the frequency domain:
\eon \int_{0}^{\infty} \lot(t) dt = \int_{0}^{1} \lot(\fto) \left(\frac{\partial \fto}{\partial t}\right)^{-1} d\fto . \eoff
We note from equation \ref{eq:fto} that
\eon \frac{\partial \fto}{\partial t} = \mto (1-\fto) + \sto \fto(1-\fto) , \eoff
\eon t(\fto)=\frac{1}{\mto+\sto} \log \left(\frac{1+(\sto/\mto)\fto}{1-\fto}\right) . \eoff
We further approximate
\begin{equation}
	\foo^* = \begin{cases}
		\moo t &\text{if } \doo \ll 2\sqrt{\mot \sot}\\
		\frac{\moo}{\doo} (1-\exp(-\doo t)) &\text{if } \doo \gg 2\sqrt{\mot \sot} .
	\end{cases}
\end{equation}
Combining these expressions and assuming $\mto \ll \sto$, we find
\eon \int_{0}^{\infty} \lot(t) dt = 2N\moo \mot \sot \gamma /\sto , \eoff
where
\begin{scriptsize}
\begin{equation}
	 \gamma = \begin{cases}
		\sto^{-1} \left(\tfrac{1}{2} \log \lp\frac{\sto}{\mto}\rp^2 - \frac{\sto}{\sot} \log (\sto/\mto) + \frac{\pi^2}{6}\right) &\text{} \doo \ll 2\sqrt{\mto \sot}\\
		\doo^{-1} \left( \log \lp\frac{\sto}{\mto} \rp - \frac{\sto}{\doo} \lp 1- \lp\frac{\sto}{\mto}\rp^{-\doo/\sto} \rp \right) &\text{} \doo \gg 2\sqrt{\mto \sot} .
	\end{cases} \label{gammaeq}
\end{equation}
\end{scriptsize}
This gives
\begin{align}
	\pcross &= 1-\exp \left[-\int \lot(t) dt\right] \notag\\
	&=1- \exp \left[\frac{-2N\moo \mot \sot \gamma}{\sto}\right] \notag\\
	&= 1- \exp \left[-2N\param\right] , \label{eq:sd_p}
\end{align}
where we have defined the useful quantity
\eon \param \equiv \frac{\moo \mot \sot \gamma}{\sto} \, . \eoff Thus we see that in very large populations, the probability of valley-crossing depends in a simple way on the single composite parameter $\Gamma$. The form of this composite parameter depends crucially on whether the fitness cost of the intermediate genotype is large or small, as defined in \eq{gammaeq}.

%----------------------------------------------------------------------------------------------------------

\section*{Discussion}

Our results have shown that the size of a population strongly influences how ``farsighted'' it can be. In small populations, genetic drift is strong relative to selection, so the evolutionary dynamics proceeds by sequential fixation. Since fixation of a deleterious intermediate becomes less likely in larger populations, this means that increasing the population size initially decreases the relative influence of fitness-valley crossing. However, as the population size increases further, beneficial mutations take longer to fix, maintaining diversity in the population and allowing double mutants to stochastically tunnel on the declining wild-type background  \cite{Iwasa2004, weinreich05b, Weissman2009}. Together, these effects lead to a non-monotonic relationship between population size and the probability that evolution will favor the complex adaptation over the directly uphill path, as illustrated in \fig{fig:data}a. This nonmonotonic dependence on population size is similar in spirit to the results of earlier work analyzing evolution on epistatic landscapes in the absence of fitness valleys \cite{Rozen2008, Handel2009, Jain2011}.

It is interesting to note that $\pcross$ does not immediately begin to rise with the onset of tunneling. Instead, the dependence is more complex, as a consequence of the tradeoff between increasing mutation rates and fixation times. Nevertheless, for populations in which the transition to valley-crossing behavior occurs in the semi-deterministic regime, we can derive a simple expression for the threshold size at which the population will tend to cross valleys with probability $\pcross$. A straightforward inversion of (\ref{eq:sd_p}) gives
\eon N= \frac{-\log \left[1-\pcross\right]}{2 \param}, \eoff
valid for large population sizes in the semi-deterministic regime. Thus in this regime the threshold size above which a population exhibits a given degree of foresight (i.e. has a particular $\pcross$) depends only on $\param$. To illustrate this, in Figure \ref{fig:data}b we show $\pcross$ as a function of $N\param$ for a variety of simulations across different values of $\mto,\moo,\mot,\sto,\doo$ and $\sot$. It is clear that even across a wide parameter range, $N \param$ is a reliable predictor of valley crossing probability in the semi-deterministic limit.

\subsubsection*{Multiple Intermediates and Evolutionary Predictability}

Recently, Szendro \emph{et al.} \cite{Szendro2013} simulated evolution across a wide range of population sizes on an experimentally-derived epistatic fitness landscape, finding that ``evolutionary entropy'' (i.e. unpredictability over all possible outcomes, given by $S = - \sum \log p_j$) varied nonmonotonically with population size. Specifically, these authors found entropy peaks at points proportional to $N \mu$ and $N \mu^2$, which they argued were related to the supply rate of single and double mutants respectively. Our analysis is consistent with these results. For example, the entropy peak at $N \mu^2$ found by Szendro \emph{et al.} \cite{Szendro2013} corresponds to our result that valley-crossing begins to significantly influence evolution when $N \sim 1/\param$: this is approximately proportional to $N \mu^2$, albeit with an additional log dependence on $\mu$ from the $\gamma$ factor that would be harder to observe experimentally.

We find related behavior if we extend our analysis to valleys with more intermediates. As a simple example, we consider a fitness landscape where we add a deleterious intermediate with fitness deficit $\doo$ to each branch, so that we now have competition between a single-intermediate valley and a two-intermediate valley. In a large enough population, mutation-selection balance will ensure that
\eon N_{u_0} = N_{i_0} = \frac{N \mu_0}{\doo} . \eoff
If we assume that these sub-populations are large enough that double-mutants behave deterministically, then we find the crossing probability obeys
\begin{small}\eon -\log \left[1-\pcross\right] = 2 \left(\frac{\mu_0}{\doo}\right) N \param = 2 \left(\frac{\sot \gamma}{\sto \doo} \right) N \mu_0 \moo  \mot . \eoff\end{small}
Thus our analysis of valleys with multiple intermediates suggests that the $N\mu^2$ entropy peak is not unique: as the population grows larger, there should be entropy peaks corresponding to foresight across valleys with increasing numbers of intermediates. The emergence of such second peaks has been observed in simulations \citep{Jain2011}, and our model offers a quantitative outline of where such peaks should occur given the relevant evolutionary parameters. In the example above, for instance, there would be an entropy peak approximately proportional to $N \mu^3$, and in general, we could expect entropy peaks at points approximately proportional to $N \mu^n$ for $(n-1)$-intermediate valleys. In practice, however, the semi-deterministic approximation will break down for any sizable number of intermediates, unless the population size is unrealistically large.

\subsubsection*{Many paths}

Throughout this paper, we have assumed the presence of a single uphill mutation and fitness valley. We now consider how our analysis can be extended to predict how evolution chooses among many such possible mutational trajectories.

In small populations that are in the sequential fixation regime, we simply add additional transient transition matrix elements representing different mutations, with the uphill mutations transitioning to the uphill absorbing state, and similarly for the valley-crossing mutations. When stochastic tunneling is important, we must instead add the rates of single valley-crossing mutants to get a total rate of
\eon \lott = \sum_v \lot,\eoff
and similarly find the total rate of uphill mutants that are destined to survive drift,
\begin{align} \ltot &= \sum_u \lto = \sum_u N \mto \pi(\sto) \notag\\&= \int  N U_b \pi(\sto) \rho(\sto) d \sto , \end{align}
where in the last equality we have replaced a discrete collection of uphill mutations with a continuous distribution of uphill fitness effects $\rho(s)$, with total beneficial mutation rate $U_b$. Once an uphill mutation destined to survive drift occurs, the probability that it has fitness $s$ is given by the ratio between its partial rate and the total rate; formally, the probability density is given by:
\eon f(s) = N \mto \ps \rho(s) / \ltot , \eoff
Using these expressions, we can integrate our results from the analysis over all possible trajectories. However, we note that if there are a large number of weakly beneficial mutations, it is possible the first successful lineage to appear will be outcompeted by a stronger uphill mutation that arises later but fixes first. Our analysis applies provided that we consider only uphill mutations that reach a significant portion $k$ of the population before a new, more fit uphill mutant is expected to be produced: i.e.
\eon \ltot \tau_k = \left(\int_{s_\text{cutoff}}^{\infty} N \mu_s \ps \rho(s) ds\right)(\tau_k) <1 .  \eoff
where $\tau_k$ is the expected time for a single-mutant destined for success to make up frequency $k$ of the population. This is consistent with our intuition that as the population size grows larger, we increasingly expect the mutations of largest effect to dominate the dynamics.

\subsection{Acknowledgments}
We thank Benjamin Good, Sergey Kryazhimskiy, Elizabeth Jerison, and other members of the Desai lab for useful discussions, and Katya Kosheleva for help with the figures. This work was supported by the Harvard HCRP, Harvard Herchel Smith, and Harvard PRISE programs (I.E.O.), and the James S. McDonnell Foundation, the Alfred P. Sloan Foundation, the Harvard Milton Fund, grant PHY 1313638 from the NSF, and grant GM104239 from the NIH (M.M.D.). Simulations in this paper were performed on the Odyssey cluster supported by the Research Computing Group at Harvard University.%\end{acknowledgments}

%\clearpage
%\newpage

%\bibliographystyle{evolution}
\bibliography{bmc_article}

\begin{thebibliography}{24}
\expandafter\ifx\csname natexlab\endcsname\relax\def\natexlab#1{#1}\fi
\expandafter\ifx\csname bibnamefont\endcsname\relax
  \def\bibnamefont#1{#1}\fi
\expandafter\ifx\csname bibfnamefont\endcsname\relax
  \def\bibfnamefont#1{#1}\fi
\expandafter\ifx\csname citenamefont\endcsname\relax
  \def\citenamefont#1{#1}\fi
\expandafter\ifx\csname url\endcsname\relax
  \def\url#1{\texttt{#1}}\fi
\expandafter\ifx\csname urlprefix\endcsname\relax\def\urlprefix{URL }\fi
\providecommand{\bibinfo}[2]{#2}
\providecommand{\eprint}[2][]{\url{#2}}

\bibitem[{\citenamefont{Perfeito et~al.}(2007)\citenamefont{Perfeito,
  Fernandes, Mota, and Gordo}}]{perfeito07}
\bibinfo{author}{\bibfnamefont{L.}~\bibnamefont{Perfeito}},
  \bibinfo{author}{\bibfnamefont{L.}~\bibnamefont{Fernandes}},
  \bibinfo{author}{\bibfnamefont{C.}~\bibnamefont{Mota}}, \bibnamefont{and}
  \bibinfo{author}{\bibfnamefont{I.}~\bibnamefont{Gordo}},
  \bibinfo{journal}{Science} \textbf{\bibinfo{volume}{317}},
  \bibinfo{pages}{813} (\bibinfo{year}{2007}).

\bibitem[{\citenamefont{Tenaillon et~al.}(2012)\citenamefont{Tenaillon,
  Rodríguez-Verdugo, Gaut, McDonald, Bennett, Long, and Gaut}}]{tenaillon12}
\bibinfo{author}{\bibfnamefont{O.}~\bibnamefont{Tenaillon}},
  \bibinfo{author}{\bibfnamefont{A.}~\bibnamefont{Rodríguez-Verdugo}},
  \bibinfo{author}{\bibfnamefont{R.~L.} \bibnamefont{Gaut}},
  \bibinfo{author}{\bibfnamefont{P.}~\bibnamefont{McDonald}},
  \bibinfo{author}{\bibfnamefont{A.~F.} \bibnamefont{Bennett}},
  \bibinfo{author}{\bibfnamefont{A.~D.} \bibnamefont{Long}}, \bibnamefont{and}
  \bibinfo{author}{\bibfnamefont{B.~S.} \bibnamefont{Gaut}},
  \bibinfo{journal}{Science} \textbf{\bibinfo{volume}{335}},
  \bibinfo{pages}{457} (\bibinfo{year}{2012}).

\bibitem[{\citenamefont{Lang et~al.}(2013)\citenamefont{Lang, Rice, Hickman,
  Sodergren, Weinstock, Botstein, and Desai}}]{lang13}
\bibinfo{author}{\bibfnamefont{G.~I.} \bibnamefont{Lang}},
  \bibinfo{author}{\bibfnamefont{D.~P.} \bibnamefont{Rice}},
  \bibinfo{author}{\bibfnamefont{M.~J.} \bibnamefont{Hickman}},
  \bibinfo{author}{\bibfnamefont{E.}~\bibnamefont{Sodergren}},
  \bibinfo{author}{\bibfnamefont{G.~M.} \bibnamefont{Weinstock}},
  \bibinfo{author}{\bibfnamefont{D.}~\bibnamefont{Botstein}}, \bibnamefont{and}
  \bibinfo{author}{\bibfnamefont{M.~M.} \bibnamefont{Desai}},
  \bibinfo{journal}{Nature} \textbf{\bibinfo{volume}{advance online
  publication}} (\bibinfo{year}{2013}).

\bibitem[{\citenamefont{Gerrish and Lenski}(1998)}]{gerrishlenski98}
\bibinfo{author}{\bibfnamefont{P.}~\bibnamefont{Gerrish}} \bibnamefont{and}
  \bibinfo{author}{\bibfnamefont{R.}~\bibnamefont{Lenski}},
  \bibinfo{journal}{Genetica} \textbf{\bibinfo{volume}{102/103}},
  \bibinfo{pages}{127} (\bibinfo{year}{1998}), \bibinfo{note}{read}.

\bibitem[{\citenamefont{Schiffels et~al.}(2011)\citenamefont{Schiffels,
  Szollosi, Mustonen, and Lassig}}]{schiffels11}
\bibinfo{author}{\bibfnamefont{S.}~\bibnamefont{Schiffels}},
  \bibinfo{author}{\bibfnamefont{G.~J.} \bibnamefont{Szollosi}},
  \bibinfo{author}{\bibfnamefont{V.}~\bibnamefont{Mustonen}}, \bibnamefont{and}
  \bibinfo{author}{\bibfnamefont{M.}~\bibnamefont{Lassig}},
  \bibinfo{journal}{Genetics} \textbf{\bibinfo{volume}{189}},
  \bibinfo{pages}{1361} (\bibinfo{year}{2011}).

\bibitem[{\citenamefont{Good et~al.}(2012)\citenamefont{Good, Rouzine, Balick,
  Hallatschek, and Desai}}]{good12}
\bibinfo{author}{\bibfnamefont{B.~H.} \bibnamefont{Good}},
  \bibinfo{author}{\bibfnamefont{I.~M.} \bibnamefont{Rouzine}},
  \bibinfo{author}{\bibfnamefont{D.~J.} \bibnamefont{Balick}},
  \bibinfo{author}{\bibfnamefont{O.}~\bibnamefont{Hallatschek}},
  \bibnamefont{and} \bibinfo{author}{\bibfnamefont{M.~M.} \bibnamefont{Desai}},
  \bibinfo{journal}{Proceedings of the National Academy of Sciences}
  \textbf{\bibinfo{volume}{109}}, \bibinfo{pages}{4950} (\bibinfo{year}{2012}).

\bibitem[{\citenamefont{Weinreich et~al.}(2005)\citenamefont{Weinreich, Watson,
  and Chao}}]{weinreich05a}
\bibinfo{author}{\bibfnamefont{D.~M.} \bibnamefont{Weinreich}},
  \bibinfo{author}{\bibfnamefont{R.~A.} \bibnamefont{Watson}},
  \bibnamefont{and} \bibinfo{author}{\bibfnamefont{L.}~\bibnamefont{Chao}},
  \bibinfo{journal}{Evolution} \textbf{\bibinfo{volume}{59}},
  \bibinfo{pages}{1165} (\bibinfo{year}{2005}).

\bibitem[{\citenamefont{Weinreich et~al.}(2006)\citenamefont{Weinreich,
  Delaney, DePristo, and Hartl}}]{weinreich06}
\bibinfo{author}{\bibfnamefont{D.~M.} \bibnamefont{Weinreich}},
  \bibinfo{author}{\bibfnamefont{N.~F.} \bibnamefont{Delaney}},
  \bibinfo{author}{\bibfnamefont{M.~A.} \bibnamefont{DePristo}},
  \bibnamefont{and} \bibinfo{author}{\bibfnamefont{D.~L.} \bibnamefont{Hartl}},
  \bibinfo{journal}{Science} \textbf{\bibinfo{volume}{312}},
  \bibinfo{pages}{111} (\bibinfo{year}{2006}).

\bibitem[{\citenamefont{Poelwijk et~al.}(2007)\citenamefont{Poelwijk, Kiviet,
  Weinreich, and Tans}}]{poelwijck07}
\bibinfo{author}{\bibfnamefont{F.~J.} \bibnamefont{Poelwijk}},
  \bibinfo{author}{\bibfnamefont{D.~J.} \bibnamefont{Kiviet}},
  \bibinfo{author}{\bibfnamefont{D.~M.} \bibnamefont{Weinreich}},
  \bibnamefont{and} \bibinfo{author}{\bibfnamefont{S.~J.} \bibnamefont{Tans}},
  \bibinfo{journal}{Nature} \textbf{\bibinfo{volume}{445}}
  (\bibinfo{year}{2007}).

\bibitem[{\citenamefont{Kvitek and Sherlock}(2011)}]{Kvitek2011}
\bibinfo{author}{\bibfnamefont{D.~J.} \bibnamefont{Kvitek}} \bibnamefont{and}
  \bibinfo{author}{\bibfnamefont{G.}~\bibnamefont{Sherlock}},
  \bibinfo{journal}{PLoS genetics} \textbf{\bibinfo{volume}{7}},
  \bibinfo{pages}{e1002056} (\bibinfo{year}{2011}).

\bibitem[{\citenamefont{Silva et~al.}(2011)\citenamefont{Silva, Mendon{\c{c}}a,
  Carvalho, Reis, Gordo, Trindade, and Dionisio}}]{Silva2011}
\bibinfo{author}{\bibfnamefont{R.~F.} \bibnamefont{Silva}},
  \bibinfo{author}{\bibfnamefont{S.~C.} \bibnamefont{Mendon{\c{c}}a}},
  \bibinfo{author}{\bibfnamefont{L.~M.} \bibnamefont{Carvalho}},
  \bibinfo{author}{\bibfnamefont{A.~M.} \bibnamefont{Reis}},
  \bibinfo{author}{\bibfnamefont{I.}~\bibnamefont{Gordo}},
  \bibinfo{author}{\bibfnamefont{S.}~\bibnamefont{Trindade}}, \bibnamefont{and}
  \bibinfo{author}{\bibfnamefont{F.}~\bibnamefont{Dionisio}},
  \bibinfo{journal}{PLoS genetics} \textbf{\bibinfo{volume}{7}},
  \bibinfo{pages}{e1002181} (\bibinfo{year}{2011}).

\bibitem[{\citenamefont{Dawid et~al.}(2010)\citenamefont{Dawid, Kiviet,
  Kogenaru, de~Vos, and Tans}}]{Dawid2010}
\bibinfo{author}{\bibfnamefont{A.}~\bibnamefont{Dawid}},
  \bibinfo{author}{\bibfnamefont{D.~J.} \bibnamefont{Kiviet}},
  \bibinfo{author}{\bibfnamefont{M.}~\bibnamefont{Kogenaru}},
  \bibinfo{author}{\bibfnamefont{M.}~\bibnamefont{de~Vos}}, \bibnamefont{and}
  \bibinfo{author}{\bibfnamefont{S.~J.} \bibnamefont{Tans}},
  \bibinfo{journal}{Chaos: An Interdisciplinary Journal of Nonlinear Science}
  \textbf{\bibinfo{volume}{20}}, \bibinfo{pages}{026105}
  (\bibinfo{year}{2010}).

\bibitem[{\citenamefont{Salverda et~al.}(2011)\citenamefont{Salverda, Dellus,
  Gorter, Debets, Van Der~Oost, Hoekstra, Tawfik, and
  de~Visser}}]{salverda2011initial}
\bibinfo{author}{\bibfnamefont{M.~L.} \bibnamefont{Salverda}},
  \bibinfo{author}{\bibfnamefont{E.}~\bibnamefont{Dellus}},
  \bibinfo{author}{\bibfnamefont{F.~A.} \bibnamefont{Gorter}},
  \bibinfo{author}{\bibfnamefont{A.~J.} \bibnamefont{Debets}},
  \bibinfo{author}{\bibfnamefont{J.}~\bibnamefont{Van Der~Oost}},
  \bibinfo{author}{\bibfnamefont{R.~F.} \bibnamefont{Hoekstra}},
  \bibinfo{author}{\bibfnamefont{D.~S.} \bibnamefont{Tawfik}},
  \bibnamefont{and} \bibinfo{author}{\bibfnamefont{J.~A.~G.}
  \bibnamefont{de~Visser}}, \bibinfo{journal}{PLoS genetics}
  \textbf{\bibinfo{volume}{7}}, \bibinfo{pages}{e1001321}
  (\bibinfo{year}{2011}).

\bibitem[{\citenamefont{Woods et~al.}(2011)\citenamefont{Woods, Barrick,
  Cooper, Shrestha, Kauth, and Lenski}}]{woods2011second}
\bibinfo{author}{\bibfnamefont{R.~J.} \bibnamefont{Woods}},
  \bibinfo{author}{\bibfnamefont{J.~E.} \bibnamefont{Barrick}},
  \bibinfo{author}{\bibfnamefont{T.~F.} \bibnamefont{Cooper}},
  \bibinfo{author}{\bibfnamefont{U.}~\bibnamefont{Shrestha}},
  \bibinfo{author}{\bibfnamefont{M.~R.} \bibnamefont{Kauth}}, \bibnamefont{and}
  \bibinfo{author}{\bibfnamefont{R.~E.} \bibnamefont{Lenski}},
  \bibinfo{journal}{Science} \textbf{\bibinfo{volume}{331}},
  \bibinfo{pages}{1433} (\bibinfo{year}{2011}).

\bibitem[{\citenamefont{Weinreich and Chao}(2005)}]{weinreich05b}
\bibinfo{author}{\bibfnamefont{D.~M.} \bibnamefont{Weinreich}}
  \bibnamefont{and} \bibinfo{author}{\bibfnamefont{L.}~\bibnamefont{Chao}},
  \bibinfo{journal}{Evolution} \textbf{\bibinfo{volume}{59}},
  \bibinfo{pages}{1175} (\bibinfo{year}{2005}).

\bibitem[{\citenamefont{Iwasa et~al.}(2004)\citenamefont{Iwasa, Michor, and
  Nowak}}]{Iwasa2004}
\bibinfo{author}{\bibfnamefont{Y.}~\bibnamefont{Iwasa}},
  \bibinfo{author}{\bibfnamefont{F.}~\bibnamefont{Michor}}, \bibnamefont{and}
  \bibinfo{author}{\bibfnamefont{M.~A.} \bibnamefont{Nowak}},
  \bibinfo{journal}{Genetics} \textbf{\bibinfo{volume}{166}},
  \bibinfo{pages}{1571} (\bibinfo{year}{2004}).

\bibitem[{\citenamefont{Weissman et~al.}(2009)\citenamefont{Weissman, Desai,
  Fisher, and Feldman}}]{Weissman2009}
\bibinfo{author}{\bibfnamefont{D.~B.} \bibnamefont{Weissman}},
  \bibinfo{author}{\bibfnamefont{M.~M.} \bibnamefont{Desai}},
  \bibinfo{author}{\bibfnamefont{D.~S.} \bibnamefont{Fisher}},
  \bibnamefont{and} \bibinfo{author}{\bibfnamefont{M.~W.}
  \bibnamefont{Feldman}}, \bibinfo{journal}{Theoretical Population Biology}
  \textbf{\bibinfo{volume}{75}}, \bibinfo{pages}{286 } (\bibinfo{year}{2009}).

\bibitem[{\citenamefont{Gokhale et~al.}(2009)\citenamefont{Gokhale, Iwasa,
  Nowak, and Traulsen}}]{gokhale2009pace}
\bibinfo{author}{\bibfnamefont{C.~S.} \bibnamefont{Gokhale}},
  \bibinfo{author}{\bibfnamefont{Y.}~\bibnamefont{Iwasa}},
  \bibinfo{author}{\bibfnamefont{M.~A.} \bibnamefont{Nowak}}, \bibnamefont{and}
  \bibinfo{author}{\bibfnamefont{A.}~\bibnamefont{Traulsen}},
  \bibinfo{journal}{Journal of theoretical biology}
  \textbf{\bibinfo{volume}{259}}, \bibinfo{pages}{613} (\bibinfo{year}{2009}).

\bibitem[{\citenamefont{Rozen et~al.}(2008)\citenamefont{Rozen, Habets, Handel,
  and de~Visser}}]{Rozen2008}
\bibinfo{author}{\bibfnamefont{D.~E.} \bibnamefont{Rozen}},
  \bibinfo{author}{\bibfnamefont{M.~G.} \bibnamefont{Habets}},
  \bibinfo{author}{\bibfnamefont{A.}~\bibnamefont{Handel}}, \bibnamefont{and}
  \bibinfo{author}{\bibfnamefont{J.~A.~G.} \bibnamefont{de~Visser}},
  \bibinfo{journal}{PLoS One} \textbf{\bibinfo{volume}{3}},
  \bibinfo{pages}{e1715} (\bibinfo{year}{2008}).

\bibitem[{\citenamefont{Handel and Rozen}(2009)}]{Handel2009}
\bibinfo{author}{\bibfnamefont{A.}~\bibnamefont{Handel}} \bibnamefont{and}
  \bibinfo{author}{\bibfnamefont{D.~E.} \bibnamefont{Rozen}},
  \bibinfo{journal}{BMC evolutionary biology} \textbf{\bibinfo{volume}{9}},
  \bibinfo{pages}{236} (\bibinfo{year}{2009}).

\bibitem[{\citenamefont{Jain et~al.}(2011)\citenamefont{Jain, Krug, and
  Park}}]{Jain2011}
\bibinfo{author}{\bibfnamefont{K.}~\bibnamefont{Jain}},
  \bibinfo{author}{\bibfnamefont{J.}~\bibnamefont{Krug}}, \bibnamefont{and}
  \bibinfo{author}{\bibfnamefont{S.-C.} \bibnamefont{Park}},
  \bibinfo{journal}{Evolution} \textbf{\bibinfo{volume}{65}},
  \bibinfo{pages}{1945} (\bibinfo{year}{2011}).

\bibitem[{\citenamefont{Van~Nimwegen and
  Crutchfield}(2000)}]{van2000metastable}
\bibinfo{author}{\bibfnamefont{E.}~\bibnamefont{Van~Nimwegen}}
  \bibnamefont{and} \bibinfo{author}{\bibfnamefont{J.~P.}
  \bibnamefont{Crutchfield}}, \bibinfo{journal}{Bulletin of mathematical
  biology} \textbf{\bibinfo{volume}{62}}, \bibinfo{pages}{799}
  (\bibinfo{year}{2000}).

\bibitem[{\citenamefont{Desai and Fisher}(2007)}]{Desai2007}
\bibinfo{author}{\bibfnamefont{M.~M.} \bibnamefont{Desai}} \bibnamefont{and}
  \bibinfo{author}{\bibfnamefont{D.~S.} \bibnamefont{Fisher}},
  \bibinfo{journal}{Genetics} \textbf{\bibinfo{volume}{176}},
  \bibinfo{pages}{1759} (\bibinfo{year}{2007}).

\bibitem[{\citenamefont{Szendro et~al.}(2013)\citenamefont{Szendro, Franke,
  de~Visser, and Krug}}]{Szendro2013}
\bibinfo{author}{\bibfnamefont{I.~G.} \bibnamefont{Szendro}},
  \bibinfo{author}{\bibfnamefont{J.}~\bibnamefont{Franke}},
  \bibinfo{author}{\bibfnamefont{J.~A.~G.} \bibnamefont{de~Visser}},
  \bibnamefont{and} \bibinfo{author}{\bibfnamefont{J.}~\bibnamefont{Krug}},
  \bibinfo{journal}{Proceedings of the National Academy of Sciences}
  \textbf{\bibinfo{volume}{110}}, \bibinfo{pages}{571} (\bibinfo{year}{2013}).

\end{thebibliography}

\clearpage
\newpage

\end{document}